\documentclass[amsmath,twocolumn,floatfix,showpacs,longbibliography]{revtex4-1}
\usepackage{natbib}

\usepackage{graphicx}

\usepackage{comment}

\begin{document}

\title{Mutual information in classical spin models}

\author{Johannes Wilms$^1$, Matthias Troyer$^2$, Frank Verstraete$^1$}
\affiliation{$^1$University of Vienna, Faculty of Physics, Boltzmanngasse 5, 1090 Wien, Austria\\%
$^2$Theoretische Physik, ETH Zurich,  8093 Zurich, Switzerland}

\begin{abstract}
The total many-body correlations present in finite temperature classical spin systems are studied using the concept of mutual information. 
As opposed to zero-temperature quantum phase transitions, the total correlations are not maximal at the phase transition, 
but reach a maximum in the high temperature paramagnetic phase.
The Shannon and Renyi mutual information in both Ising and Potts models in 2 dimensions are calculated numerically by 
combining matrix product states algorithms and Monte Carlo sampling techniques.
\end{abstract}


\maketitle

\section{Introduction}

Classical statistical mechanics models have been studied extensively
for many decades. The reason even the simple classical 2-dimensional
Ising model has managed to stay interesting for so long is due to
the non-trivial correlations arising in it, especially around the
phase transition, manifesting themselves in the divergence of thermodynamic
quantities such as heat capacity or the correlation length at the
phase transition. In this paper, we propose to study another measure
for correlations, namely, the mutual information between two parts
of a system. This quantity originates from information theory and
has been increasingly applied to problems in strongly correlated quantum many-body systems
\cite{vidal2003entanglement,calabrese2004entanglement,plenio2005entropy,its2005entanglement,kitaev2006topological,levin2006detecting,verstraete2006matrix,verstraete2006criticality,hastings2007area,fssmi,measuringrenyi,eisert2010colloquium}
but it is equally well suited to the study of classical models \cite{arealaws}, which is the subject of this paper.

The outline of this paper is as follows: In section \ref{sec:Mutual-information},
we will define mutual information and explain why it is a measure
for correlations in a system, and what properties we expect it to
have. We will proceed to show how it can be calculated
efficiently for classical spin systems, present some unexpected 
results on the location of the maximum of the mutual information (section \ref{sec:results}), and finally attempt
to give a possible physical interpretation of these results.

\section{Mutual information\label{sec:Mutual-information}}

\subsection{Definition and motivation}

The (Shannon) mutual information $I$ between any two, classical or
quantum, systems $A$ and $B$ that have possible states $a$ and
$b$ occuring with joint probability $p_{ab}$ can be defined as\[
I(A,B)=S_{A}+S_{B}-S_{AB}=\sum_{a,b}p_{ab}\log\frac{p_{ab}}{p_{a}p_{b}}\]
where $p_{a}=\sum_{b}p_{ab}$ and $p_{b}=\sum_{a}p_{ab}$ are the
marginal probabilities, obtained by summing over all the states of
the respective other system. Note how we have to sum over all the
states $(a,b)$ of the total system.

The mutual information can be understood as a sort of distance (more
precisely: Kullback\textendash{}Leibler divergence) between the actual
probability distribution of the system and a product distribution,
thereby measuring the difference to a system whose subsystems are
not correlated.

The above definition of mutual information is based on the Shannon
definition of entropy. Shannon entropies like $S_{A}=\sum p_{a}\log p_{a}$
can be seen as the limit $\kappa\to1$ of a Renyi entropy $S_{A}^{(\kappa)}=\frac{1}{1-\kappa}\log\sum_{a}p_{a}^{\kappa}$.
One good reason to study Renyi and Shannon entropies is that they
have a well defined operational meaning \cite{operationalmeaning}.
Shannon entropy provides an asymptotic description of the properties
of a system (probability distribution), in the sense that it describes
the average number of bits needed to encode a state of the system,
where the average is taken over all the states of the system, using
their respective probabilities of occurence. This is identical to
the average amount of randomness that can be extracted from the system.

The concept of mutual information quantifies the amount of information that we acquire about a part of the system A by looking at its complement B.
In contrast to a quantity like the correlation length which can be
defined using first-oder correlation functions, the concept of an
{}``order'' of a correlation does not apply to mutual information
in any of its versions. Instead, it really measures correlation in
an information-theoretic sense; it gives you the amount of information
(say, in bits) that you gain about one part of the system by looking
at the other one, thereby making use of correlations of all orders.
If entropy quantifies the amount of uncertainty in a system, $S_{A}+S_{B}-S_{AB}$
can be seen as the extra certainty, or knowledge, in the total system
as opposed to considering the systems separately -- that is the information
that connects the systems, the mutual information.

The use of mutual information as a measure of correlation can also
be understood by seeing it as the natural generalization of the entanglement
entropy, an entanglement measure for pure quantum states, to finite
temperature, i.e. mixed (Gibbs-Boltzmann) states: For pure states,
the entropy of a subsystem describes the amount of information one
part of the system has about the other, $S_{A}=S_{B}$. For mixed
states however, the entropy of the total system has to be subtracted,
and it is also possible that $S_{A}\neq S_{B}$.

To conclude this section, let us explain why we think it is relevant
to study the mutual information in classical spin models.
First of all, our intuition tells us that the total amount of correlations
in a Gibbs state should be maximized at the phase transition, as the correlation
length is diverging exactly at that point. However, as we will point out,
this notion is misguided: the mutual information is a measure not just of the
2-body but of all correlations present in the system, and this mutual
information reaches a maximum away from criticality.

A second motivation for studying mutual information in classical spin systems
is the fact that this provides a natural starting point for studying entanglement
and correlations in quantum many body systems at finite temperature;
almost no studies have been done in that respect, with the exception of the fact
that a strict area law was proven at any finite T \cite{arealaws}.
Contrasting classical to quantum behaviour is one of the main challenges in
the field of quantum information theory, and this work provides a different
approach to contrasting classical to quantum correlations.

\subsection{(Partial) partition functions}

We will now show how efficient access to {}``partial partition functions''
can help us to reformulate the formula for mutual information in such
a way that it no longer is a sum over all the states in the system
-- instead, we will just have to sum over the states of the {}``border''
between the two systems \cite{arealaws}. Of course, the sum over all states will simply
be hidden in the partition functions.

\begin{figure}
\begin{centering}
\includegraphics[width=0.8\columnwidth]{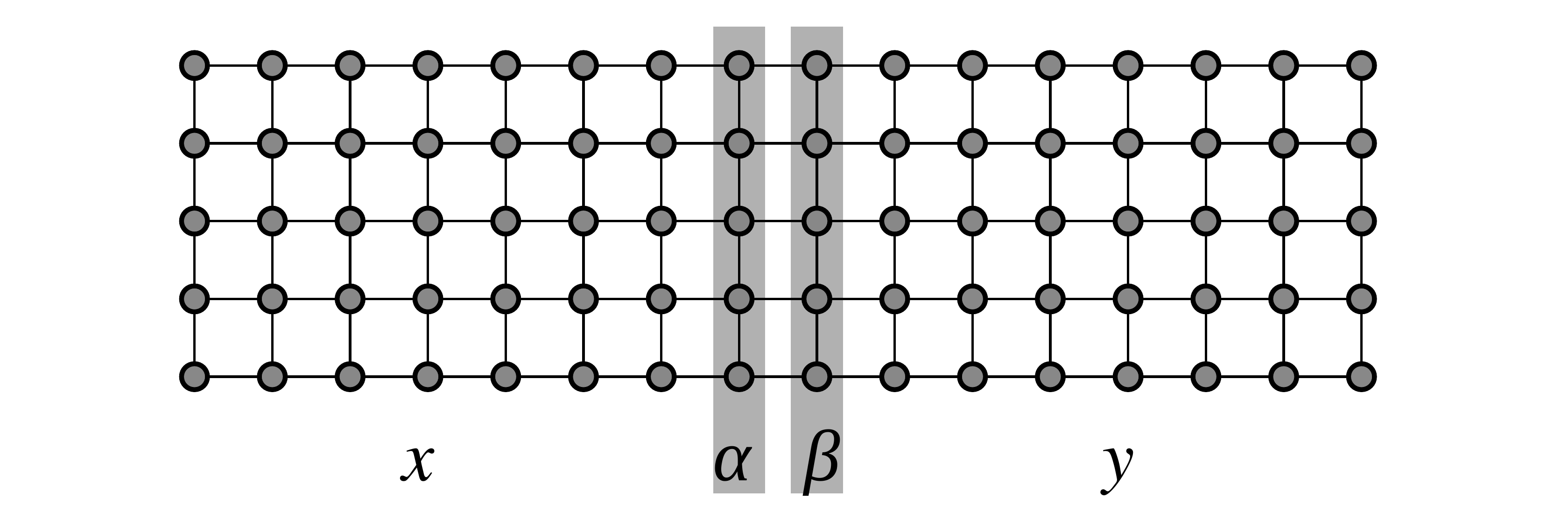}
\par\end{centering}

\caption{\label{fig:systems}Two subsystems and their interiors and boundaries}

\end{figure}

We can then divide the description of a state of the system in two
parts: the state of its {}``interior'', and the state of its {}``border''
with the other system. Here, the border is defined as the set of all
those sites that share a bond with a site that belongs to the other
subsystem. The rest of the sites makes up the interior. Figure \ref{fig:systems}
illustrates this for a simple geometry, a strip that is divided in
the middle, into {}``left'' and {}``right'' subsystems. Notice
that in this geometry both $A$ and $B$ have three sides each of
open boundaries which do not count as borders in our sense.

Let us now denote the separation into border and interior as $a=(\alpha,x)$
where $\alpha$ is the border and $x$ is the interior, and in the
same way $b=(\beta,y)$.

Then the joint probability $p_{ab}$ factors into $p_{ab}=p_{\alpha x\beta y}=p_{x}p_{\alpha x}p_{\alpha}p_{\alpha\beta}p_{\beta}p_{\beta y}p_{y}$
where $p_{x}$ describes the contribution from bonds between spins
all in the interior of $A$, $p_{\alpha x}$ the contribution from
the bonds between the interior and the border of $A$, $p_{\alpha}$
the one from bonds within the border, and so on. Plugging in this
{}``factorization'' of the probabilities, we find that we can rewrite
the Shannon mutual information as\begin{align}
I(A,B) & =\sum_{\alpha x\beta y}p_{\alpha x\beta y}\log\frac{p_{\alpha x\beta y}}{\sum_{\beta'y'}p_{\alpha x\beta'y'}\sum_{\alpha'x'}p_{\alpha'x'\beta y}}\nonumber \\
 & =\sum_{\alpha\beta}p_{\alpha}p_{\alpha\beta}p_{\beta}\underbrace{\sum_{x}p_{x}p_{\alpha x}}_{Z_{A}(\alpha)}\underbrace{\sum_{y}p_{y}p_{\beta y}}_{Z_{B}(\beta)}\nonumber \\
 & \cdot\log\frac{p_{\alpha\beta}}{\underbrace{\sum_{\beta'y'}p_{\alpha\beta'}p_{\beta'}p_{\beta'y'}p_{y'}}_{Z_{\tilde{B}}(\alpha)}\underbrace{\sum_{\alpha'x'}p_{\alpha'\beta}p_{\alpha'}p_{\alpha'x'}p_{x'}}_{Z_{\tilde{A}}(\beta)}}.\label{eq:I}\end{align}
Some explanation is required: we have identified {}``partial'' partition
functions, for example $Z_{A}(\alpha)=\sum_{x}p_{x}p_{\alpha x}$
meaning the partition function of the system $A$ with the border
$\alpha$ fixed -- we are summing over all sites $x$ in the interior
of $A$, but the border $\alpha$ of $A$ has some fixed value.

We have also defined partition functions of {}``extended'' systems.
For example, $Z_{\tilde{A}}(\beta)$ means the partition function
of the enlarged system $\tilde{A}$, which has all of $A$ (both $x$
and $\alpha)$ in its interior, and is bounded by $\beta$, which
is the fixed border configuration for this partition function.

Therefore, if we have an efficient way to calculate such partition
functions (crucially: with given fixed boundary conditions), we have
reduced the sum over all the states to a sum over just the states
of the borders.

The reduction of the mutual information to a sum that runs just over
the boundaries of the systems is analogous to the approach in \cite{arealaws}
where it was shown that this localisation of the mutual information
in the boundary implies that an area law has to be obeyed.

\subsubsection*{Renyi mutual information}

A very similar simplification as in the Shannon case also works for
Renyi mutual information $I_{AB}^{(\kappa)}=S_{A}^{(\kappa)}+S_{B}^{(\kappa)}-S_{AB}^{(\kappa)}$.
Let us introduce modified partition functions $Z_{A}^{(\kappa)}(\alpha)=\sum_{x}p_{x}^{\kappa}p_{\alpha x}^{\kappa}$
and correspondingly for the other ones. Then we calculate the Renyi
mutual information as\[
I_{AB}^{(\kappa)}=\frac{1}{1-\kappa}\log\frac{\sum_{\alpha}p_{\alpha}^{\kappa}Z_{A}^{(\kappa)}(\alpha)Z_{\tilde{B}}(\alpha)^{\kappa}\sum_{\beta}p_{\beta}^{\kappa}Z_{B}^{(\kappa)}(\beta)Z_{\tilde{A}}(\beta)^{\kappa}}{Z_{AB}^{(\kappa)}}\]
where we again only have to sum over boundaries. However, using the
Renyi mutual information does, in our models, not seem to give much
additional insight; it appears to be a smooth function of $\kappa$
around the Shannon value $\kappa=1$.

\subsubsection*{Monte Carlo}

While we will see how the partial partition functions can in many
cases indeed be evaluated efficiently, there still remains a sum over
the states of the borders in \eqref{eq:I}. While these borders between
$d-$dimensional systems are just $d-1$-dimensional, the number of
states is still exponentially large in the size of the border. While
there might be a chance to carry out the sum exactly for small system
sizes, we generally want to examine what happens towards the thermodynamical
limit. Therefore, we will employ Monte Carlo methods to sample the
exponentially large sum.

\subsection{\label{sub:strip-geometry}Strip/cylinder geometry}

The preceding section showed how the calculation of mutual information
can be simplified considerably when partial partition functions can
be evaluated efficiently. We will now show how this efficient evaluation
is possible for classical models, in the special strip-like geometry
of figure \ref{fig:systems}. In particular, let us consider a very
long strip of some given height, with a square lattice. Each column
can be considered as a transfer matrix, applied to vectors at the
(very far removed) left or right ends of the strip. Then, for an essentially
infinitely long strip, the left and right boundary conditions become
irrelevant, and each application of the transfer matrix corresponds
just to multiplying the eigenvector corresponding to the largest eigenvalue
$\Lambda$ by this largest eigenvalue.

Let us choose the two parts $A$ and $B$ simply as the left and right
half of the strip, as shown in figure \ref{fig:systems}. Let the
strip consist of $N$ columns in total, i.e. $N/2$ in each of the
subsystems, with the understanding that $N\to\infty$ such that above
approximation is justified. The partition functions are then $Z_{A}(\alpha)=\Lambda^{N/2-1}\langle\Lambda|\alpha\rangle$,
$Z_{B}(\beta)=\Lambda^{N/2-1}\langle\Lambda|\beta\rangle$, $Z_{\tilde{A}}(\beta)=\Lambda^{N/2}\langle\Lambda|\beta\rangle$,
and $Z_{\tilde{B}}(\alpha)=\Lambda^{N/2}\langle\Lambda|\alpha\rangle$.
The partition function of the whole system, which is needed for normalization
purposes, is $Z_{AB}=\Lambda^{N}.$ Let us further introduce the shorthand
$L(\alpha,\beta)=\langle\Lambda|\alpha\rangle\langle\Lambda|\beta\rangle$.

Let $q$ be the unnormalized Boltzmann weights, $q_{ab} = \exp(-\beta E_{ab})$
where $E_{ab}$ is the energy of configuration $(a,b)$. Let us now work with these
$q_{ab}$ rather than the normalized probabilities $p_{ab} = q_{ab} / \sum_{ab} q_{ab} = q_{ab} / Z_{AB}$.
The $q_{ab}$ can be factored exactly like the $p_{ab}$ in the previous section, and
we then get the following formula for the mutual information
\begin{equation}
I(A,B)=\frac{1}{\Lambda^{2}}\sum_{\alpha\beta}q_{\alpha}q_{\alpha\beta}q_{\beta}L(\alpha,\beta)\log\frac{q_{\alpha\beta}}{L(\alpha,\beta)}\label{eq:I_strip}\end{equation}
which is independent of $N$, so that the $N\to\infty$ limit is unproblematic.

\subsubsection*{Eigenvector calculation}

For this special geometry, the calculation of (partial) partition
functions essentially reduces to calculating the largest eigenvalue
of the transfer matrix, and the corresponding eigenvector. The transfer
matrix has a very special form: it is a matrix product operator (MPO).

The eigenvalues of an MPO can be calculated efficiently by a variational
algorithm \cite{nishino1,nishino2,mpsreview,pbcmps}, and in fact calculating
the extremal eigenvalues is particularly simple. The algorithm produces
the eigenvector in the form of a matrix product state (MPS), which
is a much more compact description than the full exponentially large
vector would be. There is also no problem if we decide to connect
the top and bottom of the strip by periodical boundary conditions,
thereby turning the strip into a cylinder.

Of course, in any case the MPS is only an approximation, with the
quality of the approximation depending on the chosen virtual bond
dimension of the MPS. But for the Ising transfer matrix it turns out
that we can get a reasonably good approximation with very low bond
dimension like $D=8$ (of course, to completely avoid any errors,
the bond dimension would have to grow exponentially with the system
size; but we find that larger bond dimensions do not significantly
change our results any more, so the small bond dimensions seem to
catch the essential features of our system).

There is still an exponentially large sum in \eqref{eq:I_strip} though,
which we cannot get rid of in principle, but which can be simplified
as shown in the following:

\subsubsection*{\label{sub:strip-simplifications}Simplifications}

We note that \eqref{eq:I_strip} would be a lot simpler if the logarithmic
term wasn't in there, because then we would just have

\begin{equation}
\sum_{\alpha\beta}\langle\Lambda|\alpha\rangle q_{\alpha}q_{\alpha\beta}q_{\beta}\langle\beta|\Lambda\rangle=\langle\Lambda|TT|\Lambda\rangle\label{eq:simplesumTT}\end{equation}
with $T$ the Ising transfer operator, and because that can be written
as an MPO, this expression would be easy to calculate.

Of course we cannot simply drop the logarithmic term; but we can separate
the logarithmic term into three parts, $\log\left(q_{\alpha\beta}/\left(\langle\Lambda|\alpha\rangle\langle\Lambda|\beta\rangle\right)\right)=\log q_{\alpha\beta}-\log\langle\Lambda|\alpha\rangle-\log\langle\Lambda|\beta\rangle$.

Now, $q_{ab}$ is actually an exponential, of the sum over all the
bonds between two columns: $q_{ab}=\exp(-K\sum_{i}\alpha_{i}\beta_{i})$,
where the sum goes over the rows $i$ and $\alpha_{i}\in\{-1,+1\}$
and $\beta_{i}\in\{-1,+1\}$ are the components of the configurations
$\alpha$ and $\beta,$ respectively.

The logarithm of the exponential is of course just the exponent $-K\sum_{i}\alpha_{i}\beta_{i}$,
so we have to calculate\begin{equation}
-K\sum_{\alpha\beta}\langle\Lambda|\alpha\rangle q_{\alpha}q_{\alpha\beta}q_{\beta}\langle\beta|\Lambda\rangle\sum_{i}\alpha_{i}\beta_{i}.\label{eq:simplesum1}\end{equation}
We can now consider the terms separately for each $i$, and notice
that they are all local and can therefore efficiently be calculated
as a contraction with suitably modified MPOs.

What about the remaining parts, with $\log\langle\Lambda|\alpha\rangle$
and $\log\langle\Lambda|\beta\rangle$? Let us look only at the one
with $\log\langle\Lambda|\alpha\rangle$; the one with $\log\langle\Lambda|\beta\rangle$
is exactly the same, due to symmetry. We have\[
\begin{aligned} & -K\sum_{\alpha\beta}q_{\alpha}q_{\alpha\beta}q_{\beta}\langle\Lambda|\alpha\rangle\langle\Lambda|\beta\rangle\log\langle\Lambda|\alpha\rangle\\
= & -K\sum_{\alpha}q_{\alpha}\underbrace{\sum_{\beta}q_{\alpha\beta}q_{\beta}\langle\Lambda|\beta\rangle}_{\Lambda\langle\Lambda|\alpha\rangle}\langle\Lambda|\alpha\rangle\log\langle\Lambda|\alpha\rangle\\
= & -K\Lambda\sum_{\alpha}q_{\alpha}\langle\Lambda|\alpha\rangle^{2}\log\langle\Lambda|\alpha\rangle\\
= & -\frac{1}{2}K\Lambda\sum_{\alpha}q_{\alpha}\langle\Lambda|\alpha\rangle^{2}\log\langle\Lambda|\alpha\rangle^{2}\\
= & -\frac{1}{2}K\Lambda\sum_{\alpha}q_{\alpha}\langle\Lambda|\alpha\rangle^{2}\log q_{\alpha}\langle\Lambda|\alpha\rangle^{2}\\
 & +\frac{1}{2}K\Lambda\sum_{\alpha}q_{\alpha}\langle\Lambda|\alpha\rangle^{2}\log q_{\alpha}\end{aligned}
\]
where we introduced a $q_{\alpha}/q_{\alpha}$ unit term in the logarithm
and used it to separate into two parts, one that has the form of an
entropy, $H=-\sum_{\alpha}\pi_{\alpha}\log\pi_{\alpha}$, and one
additional term where we again have the logarithm of an exponential,
which becomes a sum of local terms that can be handled easily.

So what remains is just the entropy term that describes the entropy
$-\sum_{\alpha}\pi_{\alpha}\log\pi_{\alpha}$ of the marginal distribution
$\pi_{\alpha}=q_{\alpha}\langle\Lambda|\alpha\rangle^{2}$, normalized
by $\sum_{\alpha}\pi_{\alpha}=\Lambda$. If we have the eigenvector
$|\Lambda\rangle$, we can calculate this entropy: for small numbers
of rows we can do it by exact summing, and for large numbers of rows
it can be approximated by Monte Carlo sampling. In particular, note
that for sampling the coefficients $\langle\Lambda|\alpha\rangle$
it is sufficient to know the eigenvector as a MPS, without ever needing
it in exponentially large full form. We can do the sampling particularly
efficiently by using Monte Carlo updates that sweep back and forth
along the MPS and store intermediate contraction results.

\subsection{Nested rectangular geometry}

The MPS formalism is very well suited for the strip- or cylinder-like
geometry described before, and it has the huge advantage that it can
just as well be used for not exactly solvable models. However, there
is also a different kind of geometry that is another very natural
candidate to be examined for the behaviour of the mutual information,
namely, a nested geometry, where one of the systems is contained entirely
within the other (in fact, this was the first system we decided to
study). This geometry is shown in the small left inset of figure \ref{fig:rect}.
While the MPS approach is not in principle unsuitable for this geometry,
we still opted to use different techniques instead, which also has
the advantage of serving as an independent way of verification of
results. Unlike the MPS method, these kind of techniques will not
work for all classical models, but only for exactly solvable ones.

\subsubsection*{The Fisher-Kasteleyn-Temperley method }

This method can be formulated without any understanding of the (partial)
partition functions as tensor network contractions. As detailed in
for example \cite{fisher1966dspim,mccoy1973tdim}, the Ising partition
function can be expressed as a sum of weighted dimer coverings on
an extended orientable lattice. Handling of the fixed boundary conditions
is the only essential new ingredient in our use of the FKT method,
and it may be worth discussing it shortly: Being able to fix the directions
of arbitrary spins in the lattice would be very similar to being able
to solve the model in the presence of magnetic fields, which is not
possible using this method. However, we only want to fix spins on the boundary
of a system. Our approach to do this is to simply connect all these
spins with bonds of infinite strength (this can in fact be done without
introducing actual infinities). That way, their relative orientation
is fixed, and all that remains is a factor of 2 in the partition function
describing how the whole boundary can be flipped.

This sum of weighted dimer coverings can then be calculated as the
Pfaffian of a matrix, the dimension of this matrix scaling linearly
in the number of sites in the system. The Pfaffian can be easily calculated
as the square root of a determinant, therefore with polynomial complexity.

\section{Results\label{sec:results}}

We examined the behaviour of the mutual information as a function
of the system parameters in the classical 2D Ising model on a square
lattice, defined by the Hamiltonian $H=-J\sum_{\langle ij\rangle}s_{i}s_{j}$
where the sum is over all nearest neighbours and the $s_{k}$ take
the values $\pm1$. One of the reasons for choosing the classical
Ising model was the fact that there is just the one parameter $K=J/k_{B}T$
describing the bond strength in units of $k_{B}T$. We are however
simulating finite systems, so clearly the system size, or more exactly
sizes, are another group of parameters.

Let us first consider a system of some given fixed size though. How
should the mutual information behave as a function of temperature?
Clearly, for high temperatures (small $K$) the mutual information
should tend to zero -- looking at one subsystem will not reveal anything
about the other one, as all the spins are oriented randomly.

What happens for low temperatures? Again, the situation is easy to
understand in our choice of the Ising model: There is a degenerate
ground state manifold, spanned by the two states with either all spins
pointing up or all spins pointing down. Therefore, at zero temperature
(large $K$) the mutual information should be exactly one bit: If
we see that one system has its spins pointing up, then we know that
the spins in the other system will also point up, rather than down,
and this is exactly the answer to one question with a binary answer.

What will happen at intermediate temperatures? The Ising model undergoes
a phase transition at $K\approx0.4407$, from the paramagnetic to
the ferromagnetic phase. At a phase transition, the correlation length
diverges in the thermodynamic limit. The mutual information describes
correlations between the two subsystems. It is therefore natural to
assume that the mutual information should have a maximum at the critical
point, in a similar fashion like the entanglement entropy of ground
states of several quantum models has been found to have a maximum
(that actually becomes a singularity) at the quantum critical point
\cite{eentLMG,eentcoll}. However, in our case the mutual information
does not have a maximum at the critical temperature:

\subsection{Strip/cylinder geometry}

\begin{figure}
\includegraphics[width=1\columnwidth]{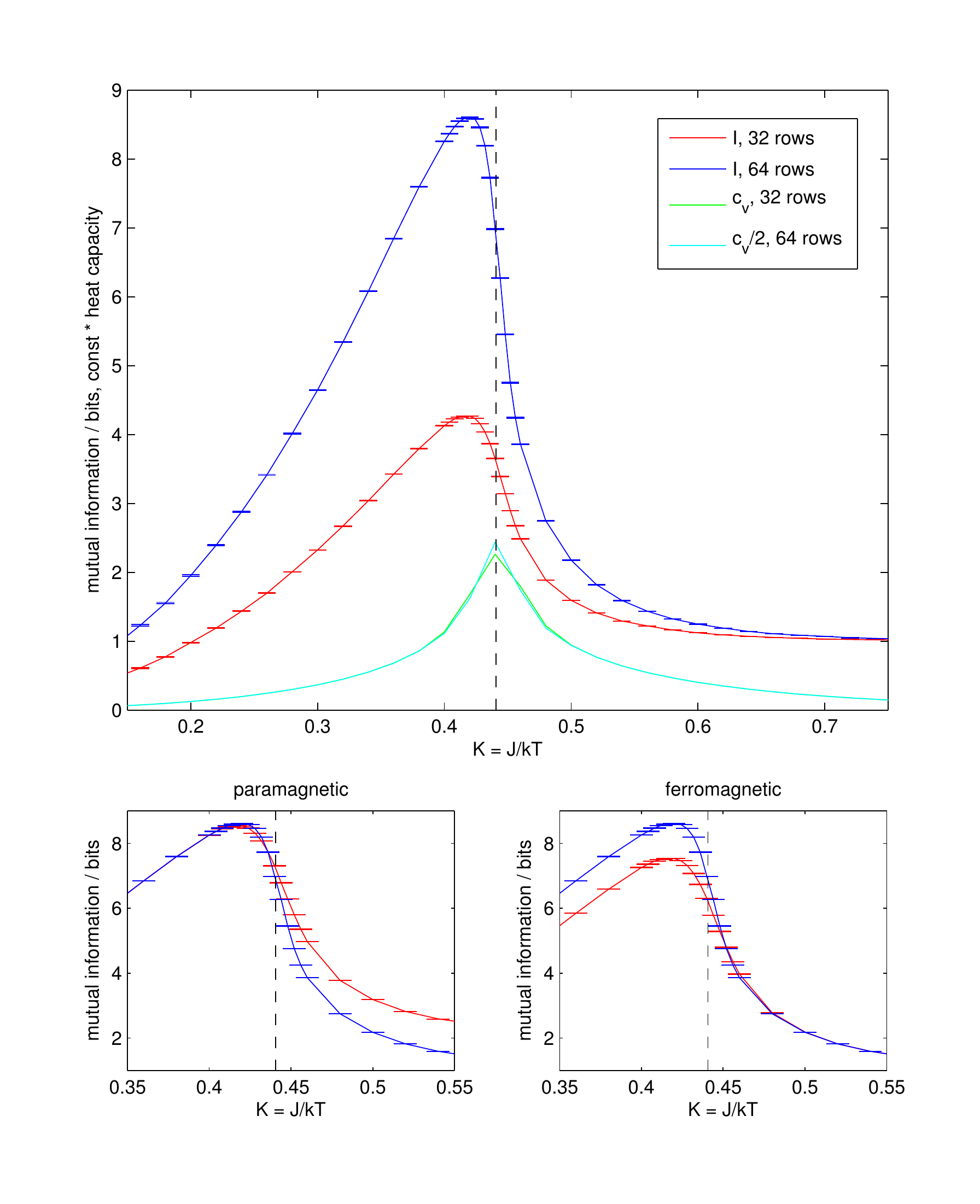}

\caption{\label{fig:pbc}Mutual information for the PBC strip geometry}

\end{figure}

Figure \ref{fig:pbc} shows the mutual information as a function of
temperature, for the cylinder-like geometry, i.e. a strip with periodic
boundary conditions in the finite dimension, with either 32 or 64
sites in that direction. It was calculated as described in section
\ref{sub:strip-geometry}, using Monte Carlo sampling. The error bars
due to Monte Carlo sampling are indicated.

We can see the predicted low- and high-temperature behaviour. There
also is a maximum, as expected. However, the maximum is not at the
critical point. It wouldn't be unreasonable to first assume that this
is due to considering a finite system as opposed to the thermodynamic
limit. The first indication that this is not the case is given by
also plotting the heat capacity (up to constant factors), which is
just a suitable derivative of the total partition function. The maximum
of the heat capacity matches very well the critical point, thereby
suggesting that the system in question is already a reasonably good
approximation to the thermodynamic limit.

Also, once the system size is chosen suffiently large, the location
of the maximum changes only very slightly with system size. It then
lies deeply within the paramagnetic phase, and increasing the system
size only appears to move it deeper into it. This is shown the lower
part of figure \ref{fig:pbc}, where in the bottom left we just multiplied
the data for 32 rows with a factor of two and got very good agreement
in the paramagnetic (high-temperature) phase, so clearly the maximum
is not affected much by doubling the system size. In the bottom right
the data for 32 rows is multiplied by a factor of two as well, but
we subtracted 1 to get agreement in the limit of very low temperatures.
It turns out we again find very good agreement for the whole ferromagnetic
phase, right up to the critical point \emph{-- not} the maximum of
the mutual information.

The bond dimension of the MPS approximation used for the data in the
picture was just 8, which was chosen such that the dominant source
of error is the Monte Carlo sampling, not the MPS approximation --
choosing a smaller bond dimension and thereby allowing more samples
to be taken in the same time appears the preferable approach. Only
in the immediate vicinity of the phase transition with the diverging
slope the error due to the MPS approximation is of the same order
of magnitude as the sampling error (and the error bars should therefore
no longer be taken too literally there) -- however, this does not
affect the location of the maximum in any way.

It should be noted that in order to reproduce our results for low
temperatures, the eigenvector given by the MPS routines has to be
suitably symmetrized. There, the ground state is almost degenerate,
and instead of the actual ground state the MPS algorithms tend to
find a different state within that approximate ground state manifold.

The fact that the curves for 32 and 64 rows agree so nicely does not
mean that there is no influence of system size. For smaller system
sizes, this is clearly the case. For these smaller system sizes, we
can even work without Monte Carlo sampling, thus eliminating this
as a possible source of error. We can also even use the exact eigenvector
as opposed to its MPS approximation. We found that in fact increasing
the system size moves the maximum montonously deeper into the paramagnetic
phase. The location of the maximum appears to move towards a value
around $K\approx0.41$, although it we could not fit any reasonable
function and therefore cannot even see if there is convergence to
any definite value at all. Quite as expected though, systems with
periodic boundary conditions behave like much bigger systems with
open boundary conditions.

It should be pointed out that the known location of the phase transition
still appears to mark a relevant position in the plot: Just as has
been found in \cite{fssmi} for different (quantum) models, it seems
to be an inflection point; and in fact -- at least in our case --
one where the first derivative tends to minus infinity in the thermodynamic
limit (as follows from the different scaling behaviours in the different
phases).

\subsection{Nested rectangular geometry}

\begin{figure}
\includegraphics[width=1\columnwidth]{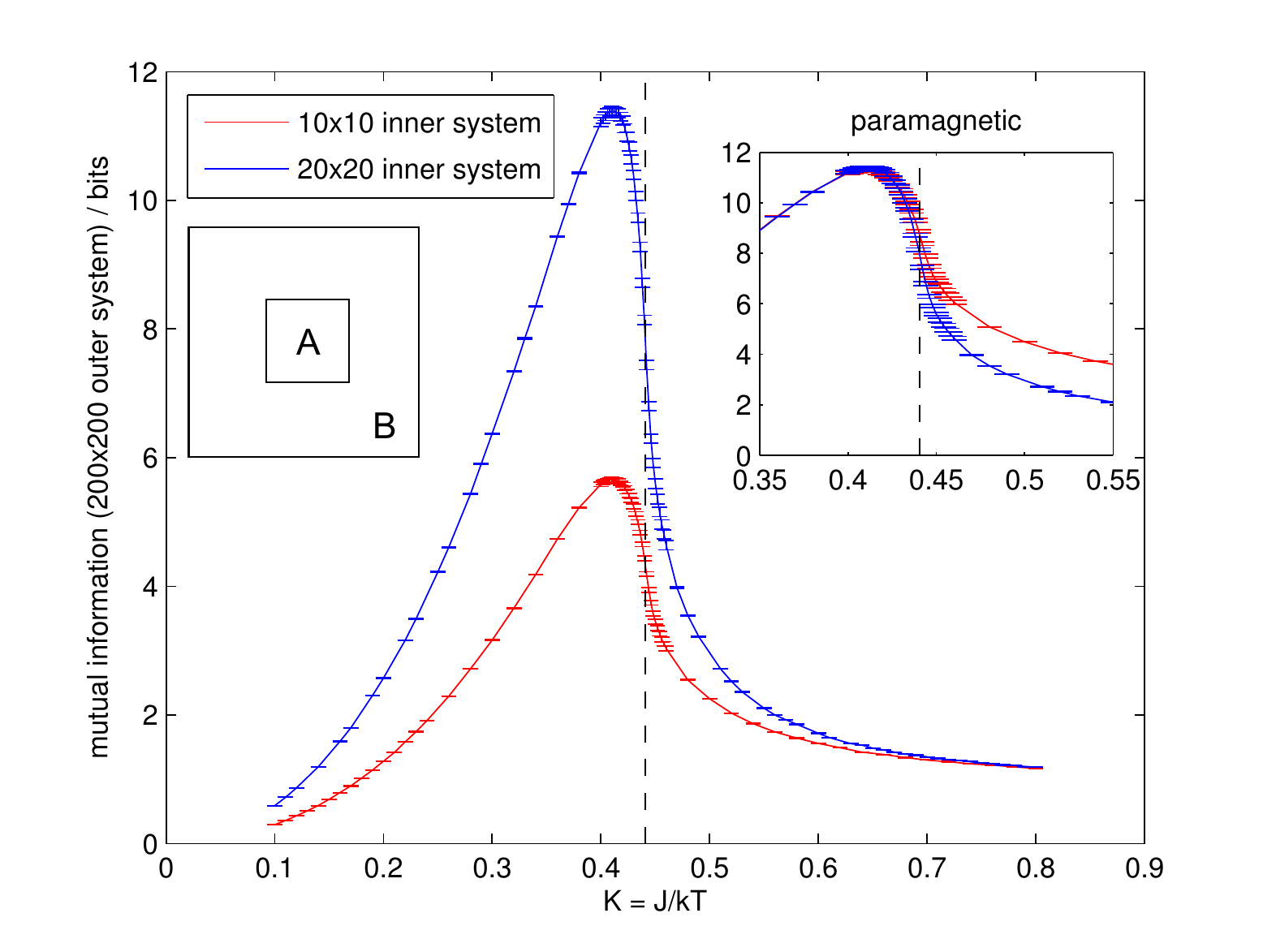}\caption{\label{fig:rect}Mutual information for the nested rectangular geometry}

\end{figure}

The results for the nested rectangular geometry are mainly presented
to reinforce that the data shown in the previous section do not appear
to be an anomaly, but describe the actual physics. Figure \ref{fig:rect}
presents data where we have a {}``square inside a square'' geometry,
with the outer square chosen so big as to hopefully avoid effects
due to its size. This is confirmed by the fact that the data for an
outer system with just $100\times100$ sites looks virtually identical,
except for some slight finite-size effects around criticality.

Again, the maximum of the mutual information is clearly located in
the paramagnetic phase, and in that phase it is again possible to
{}``collapse'' the data by multiplying the mutual information of
the smaller system with a factor of two. The ferromagnetic phase appears
to be somewhat more complicated in this geometry.

\section{Other models}

\begin{figure}
\includegraphics[width=1\columnwidth]{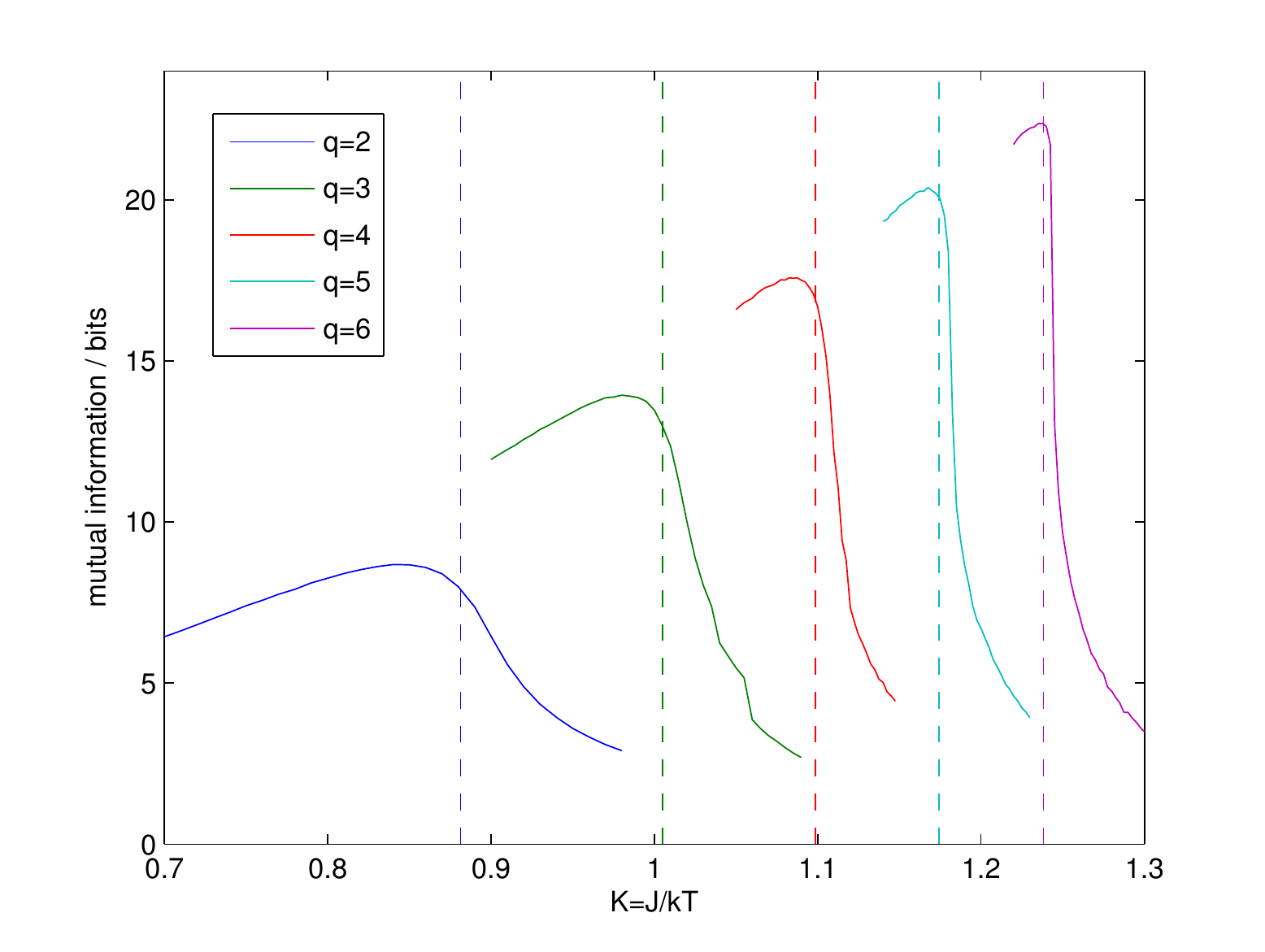}

\caption{\label{fig:potts-q}Potts models with different values of $q$. Strip
geometry (with open boundary conditions), 64 rows, MPS bond dimension
$D=16$}

\end{figure}

Let us take a brief look at a few other models: Figure \ref{fig:potts-q}
shows plots of the mutual information in Potts models, which are defined
by the Hamiltonian $H=-J\sum_{\langle ij\rangle}\delta(s_{i},s_{j})$,
where the $s_{k}$ can now however take $q$ different values instead
of just two as in the Ising model. The $q=2$ case is of course identical
to the Ising case, apart from a scaling of the coupling constant by
a factor of two. Also indicated in the plot are the exactly known
critical temperatures. Nothing much changes qualitatively by increasing
$q$, but the maxima become narrower and closer to the exact phase
transition. It might be noted that for $q\leq4$ the phase transition
is a continuous phase transition, while for $q>4$ it is of first
order, but this does not appear to reflect in any way in the behaviour
of the mutual information.

\begin{figure}
\includegraphics[width=1\columnwidth]{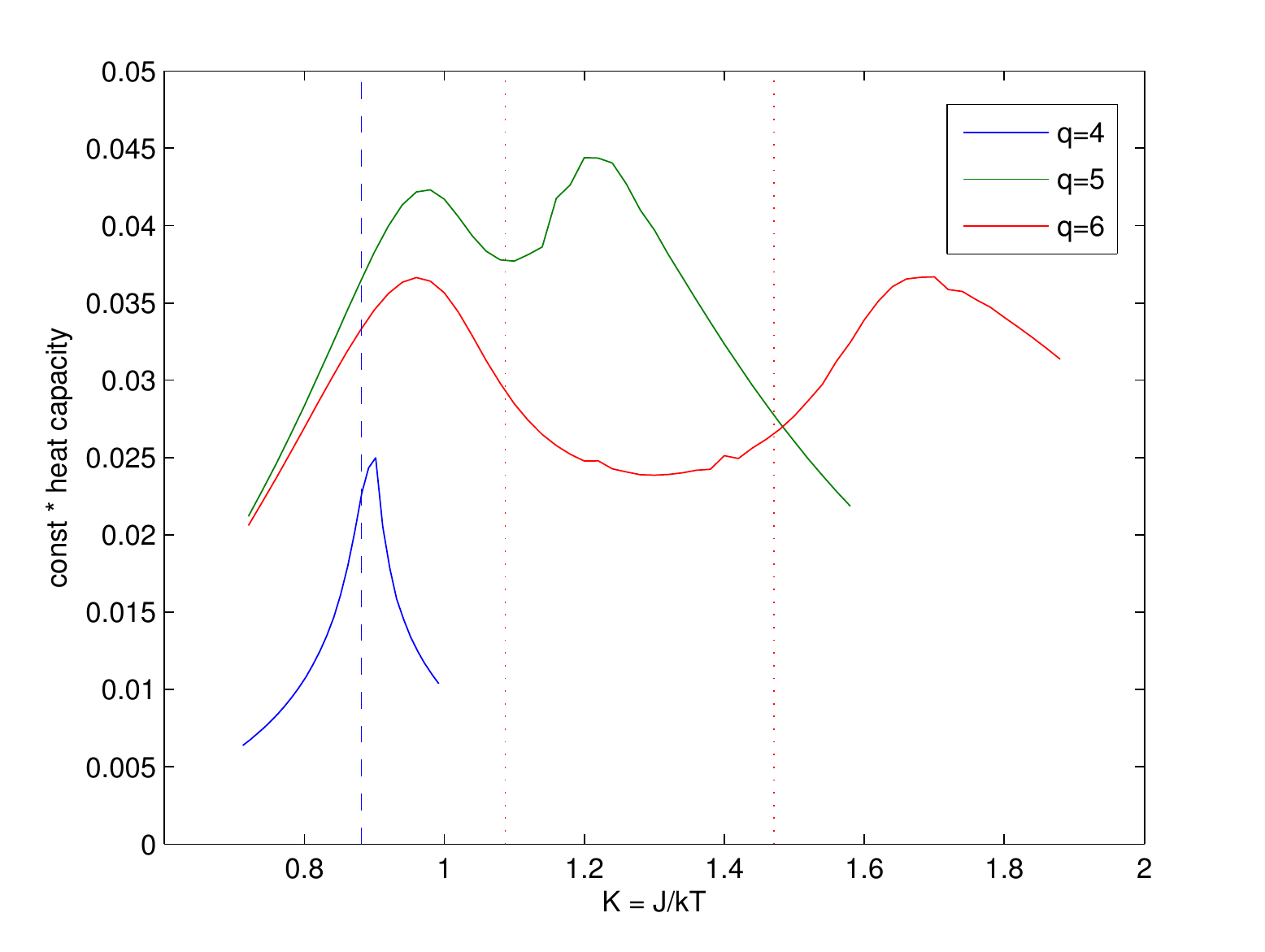}\caption{\label{fig:clock-hc}Heat capacity for the clock model with $q=4,5,6$}

\end{figure}

\begin{figure}
\includegraphics[width=1\columnwidth]{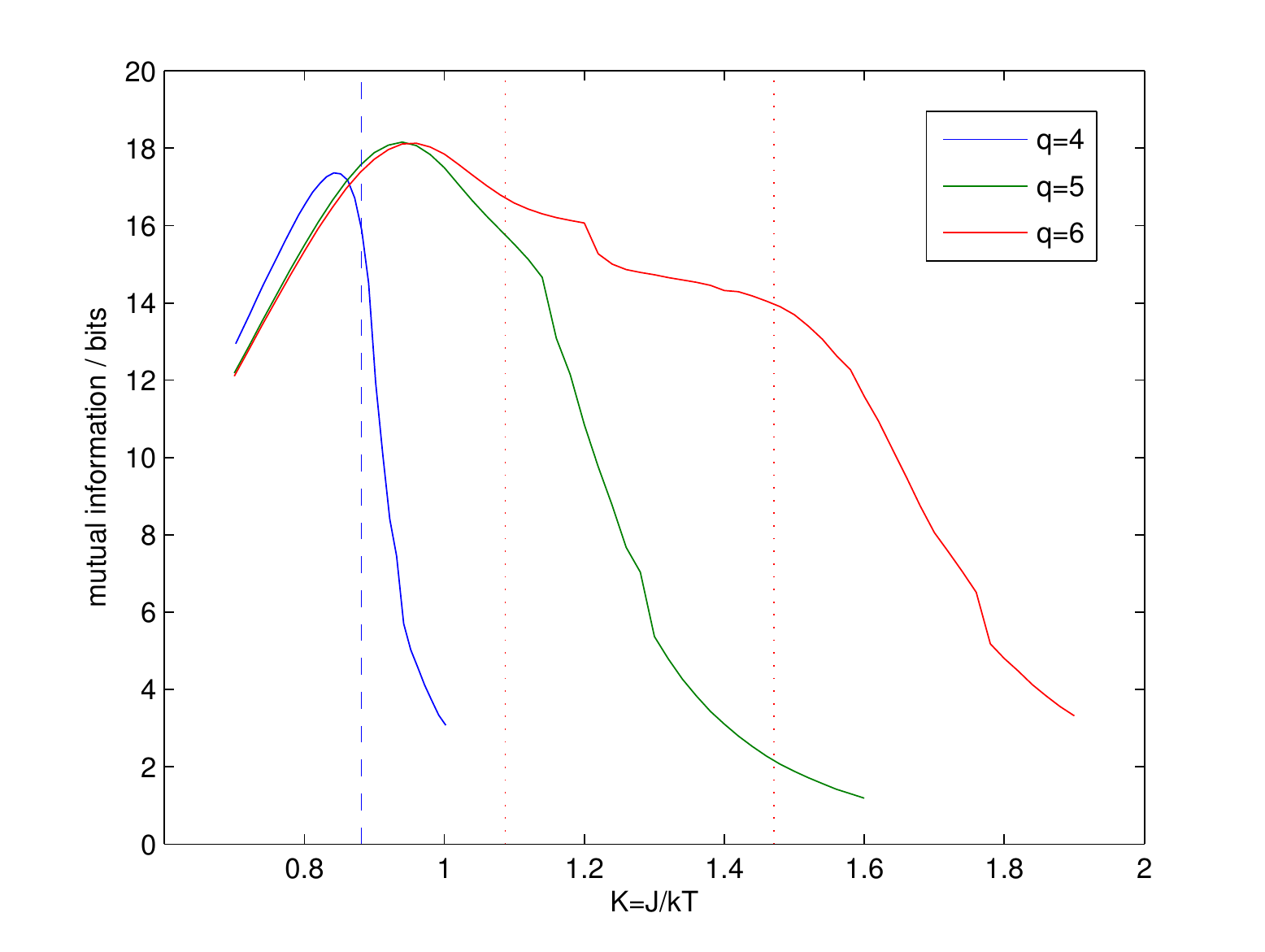}\caption{\label{fig:clock-mi}Heat capacity for the clock model with $q=4,5,6$}

\end{figure}

A different generalization of Ising models is given by Potts models,
governed by the Hamiltonian $H=-J\sum_{\langle ij\rangle}\cos(s_{i},s_{j})$
where the $s_{k}$ are now angles $2\pi k/q$. For higher $q$, this
is in several ways a more complex model, which is why figure \ref{fig:clock-hc}
first shows ours numerical results for the heat capacity. The behaviour
of the heat capacity matches well that found in \cite{clock456,clock456erratum,clock6}.
It indicates that the clock model does in fact have two phase transitions
for $q>4$. However, these two critical points no longer coincide
well with the maxima of the heat capacity; the dotted red lines show
the locations of the phase transitions as estimated in \cite{clock6}.
For the $q=4$ case, the maximum of the heat capacity is a much better
match for the location of the phase transition, which is known exactly
in this case \cite{clock4}. Figure \ref{fig:clock-mi} then shows
the corresponding mutual information plots. The $q=4$ case follows
very much the usual pattern we have found before, with the phase transition
occurring at the inflection point of the mutual information, but the
meaning of the curves for higher $q$ is essentially open for interpretation.

\section{Relationship to Fortuin-Kasteleyn clusters  \label{sec:clusters}}

Is there a way to understand why, in all the cases we have studied,
the mutual information should have its maximum not at the phase transition,
but in the high-temperature phase? We would like to suggest the idea
to look at the Fortuin-Kasteleyn (FK) clusters \cite{fortuinkasteleyn}
in the model. Those clusters are clusters of aligned spins, but not
the simple geometric clusters that would arise from combining all
neighbouring aligned spins. Instead, bonds between aligned spins only
exist with a probability $1-\exp(\beta\Delta E)$ where $\Delta E$
is the energy difference between aligned and unaligned spins. Making
bonds with this probability ensures that we only use {}``meaningful''
bonds and not those existing just because two neighbouring spins happen
to be aligned. The resulting FK clusters can then be flipped independently
from each other, which is the basis for efficient Monte Carlo cluster
updates.
In any given configuration during the Monte Carlo process, spins within a
cluster are perfectly correlated, while spins in different clusters are uncorrelated.

\begin{figure}
\includegraphics[width=1\columnwidth]{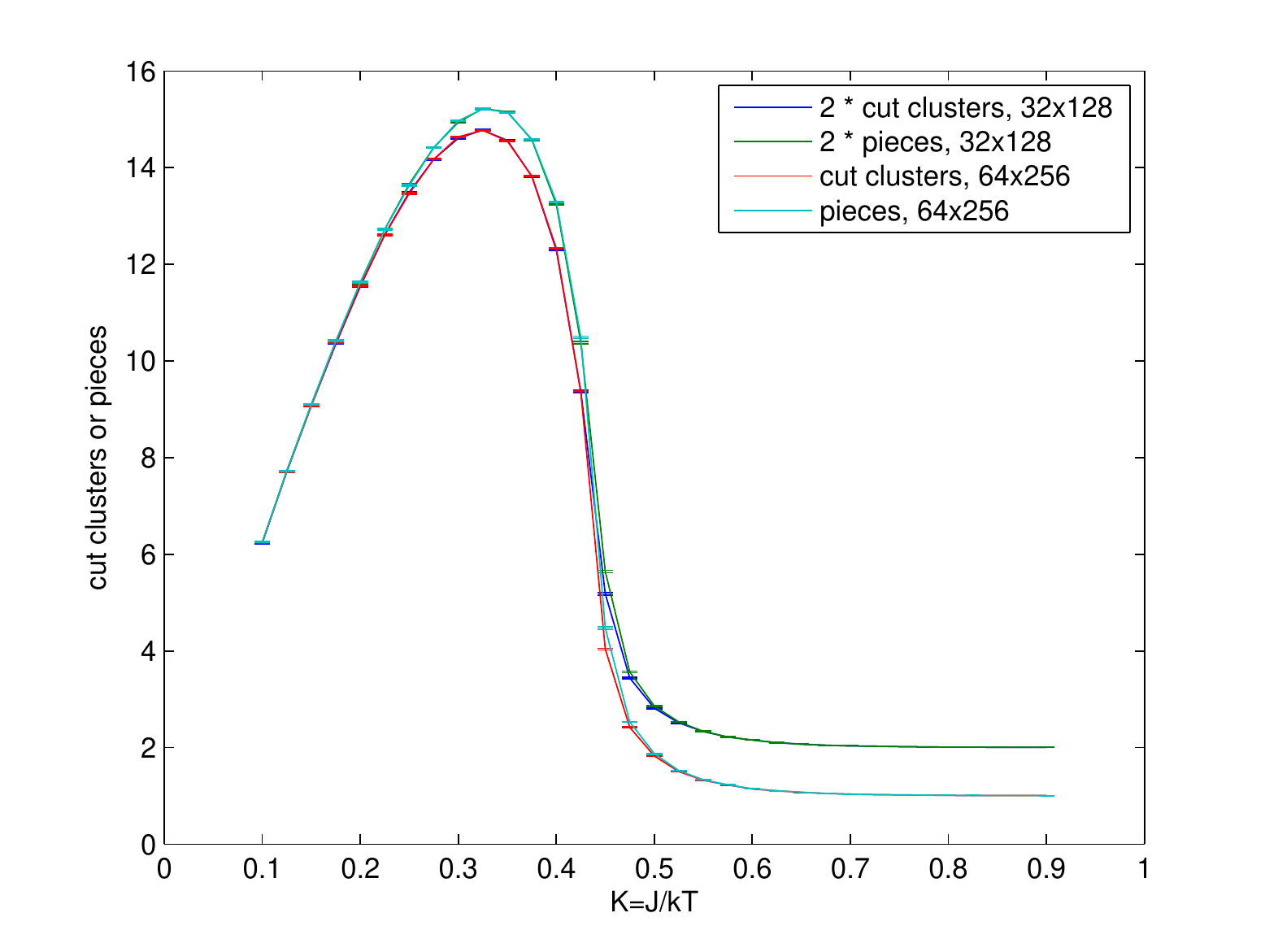}\caption{\label{fig:clusters}FK clusters being cut (cylindrical geometry).
The simulations were done using Swendsen-Wang \cite{swendsenwang}
cluster updates, which also allow the immediate identification of
clusters that are being cut. Code from the ALPS project \cite{ALPS,ALPSscheduler}
was used in the simulations.}

\end{figure}

While we cannot currently state an exact relation,
it seems intuitive that the mutual information should be related to
the number of such clusters that are cut when dividing the total system
into its two subsystems. Possibly, an even better quantity to study
would be the number of pieces that result from such a cut. The mutual
information is certainly something more complicated than just the
number of those pieces, but it might nevertheless help to give an
insight of the factors at play: at very high temperature, FK clusters
(unlike geometric clusters!) only consist of single spins, therefore
no clusters will ever be cut, corresponding to no mutual information.
At very low temperatures, there will just be one big cluster, corresponding
to the limit of one bit (in Potts models: $q$-it) of mutual information.
In between, there will be a maximum, where we have several clusters
of nontrivial size that are being cut. Figure \ref{fig:clusters}
shows this for the simple measures {}``number of clusters cut''
or {}``number of pieces (on one side) after the cut'', for the Ising
model. While it is obvious that these measures do not show the same
behaviour as the mutual information, it seems notable that they both
also exhibit maxima that lie deeply in the paramagnetic phase, and
there again appears to be a similar inflection point at the critical
temperature. This could serve as a starting point for understanding
this phenomenon.%

\section{Conclusion}

We have presented methods that allow to calculate mutual information
in classical models. We examined the scaling behaviour in different
phases and concluded that the system sizes we can study are a good
approximation of the thermodynamic limit. We still would like to gain
a better understanding of the maximum of the mutual information within
the high-temperature phase. We are also still studying different models
and anisotropic couplings, and are also working on extending our results
to quantum rather than just classical models. For example, the
$1+1$-dimensional quantum Ising model is of course intimately related to the
$2$-dimensional classical Ising model studied in the present work, by a
mapping that has also been exploited in \cite{critentropy}.

\begin{acknowledgments}
We thank Hans Gerd Evertz, Maarten van den Nest, and Gerardo Ortiz for many valuable comments and discussions.
This work is supported by the EU STReP QUE-VADIS,
the ERC grant QUERG, the FWF SFB grants FoQuS and
ViCoM, the FWF Doctoral Programme CoQuS (W 1210) and the Pauli Center at ETH Zurich.
\end{acknowledgments}

\bibliographystyle{apsrev4-1}
\bibliography{refs}

\end{document}